\documentclass[fleqn,twoside,dvips]{mystyle}
\usepackage{fancyheadings}
\usepackage[tbtags]{amsmath}
\usepackage{amssymb}
\usepackage{graphicx}
\usepackage{fouridx}
\usepackage{cite}
\usepackage{txfonts}
\def\bm#1{\mbox{\boldmath $#1$}}

\newcommand{\etal}{{\it et al.}}
\newcommand{\NN}{\nonumber \\}

\def\br{{\bm{r}}}
\def\bp{{\bm{p}}}

\newcommand{\Nuc}[4]{\fourIdx{#2}{#3}{}{#4}{\rm #1}}
\newcommand{\nuc}[2]{\mbox{$^{#2}${#1}}}
\def\Re{\mathop{\rm Re}}

\def\<{\langle}
\def\>{\rangle}

\begin{document}
\title{Semiclassical origin of asymmetric nuclear fission:
Nascent-fragment shell effect in periodic-orbit theory}

\author{
{\large\textbf{\textsf{Ken-ichiro Arita,$^1$ Takatoshi Ichikawa$^2$
and Kenichi Matsuyanagi$^{2,3}$}}} \\
$^1${\rm Department of Physics, Nagoya Institute of Technology,
Nagoya 466-8555, Japan} \\
$^2${\rm Yukawa Institute for Theoretical Physics, Kyoto 606-8502, Japan} \\
$^3${\rm RIKEN Nishina Center, Wako 351-0198, Japan}
}


\date{Submitted: 28 June 2019\\ Revised: 20 August 2019}

\iopabs{
The origin of the asymmetry in the fragment mass distribution of
low-energy nuclear fission is considered from the semiclassical point
of view.  Using the semiclassical periodic-orbit theory, one can
define and quantify the shell effect associated with spatially
localized nascent-fragment (prefragment) part of the potential.  We
investigate the roles of prefragments in the deformed shell effect
using a simple cavity potential model, but with realistic shape
degrees of freedom for describing the fission processes.  The results
suggest that the prefragment magic numbers play essential roles in
determining the shapes at the fission saddles, which should have a
close relation to the fragment mass distribution.
}

\maketitle

\section{Introduction}
\label{sec:intro}

In the low-energy fission process of a heavy nucleus, nucleon
distribution is elongated in one direction and a neck is formed which
begins to separate the system into two nascent fragments, which we
shall call ``prefragments'' for shortness.  By the Coulomb repulsion
between two prefragments, the system is finally divided into two
fragments.  According to the experimental results\cite{Andr}, the
fragment mass distributions are asymmetric in most of the actinide
nuclei, namely, those nuclei are likely to break up into two fragments
with different sizes.  Since the fragment mass distribution is
determined by the shape of the nucleus on the fission saddle, the
system is expected to favor an asymmetric shape in the fission
deformation processes.  The fission process is first studied with the
liquid-drop model (LDM)\cite{BW,HW}.  However, the asymmetric
fragment-mass distribution cannot be explained within the LDM:
symmetric shapes are favored throughout the fission deformation
processes (see Sec.~\ref{sec:tqs} below).  The above problem is known
to be solved by taking account of the quantum shell effect.  Both
static and dynamic theoretical approaches have achieved great
successes in the systematic reproduction of the experimental
results\cite{Schmidt}.

The most remarkable feature of the experimental results in the
fissions of actinide nuclei would be the preference of heavier
fragments around $A\sim 140$ regardless of the parent species.  It was
considered as due to the strong shell effect of the spherical
fragments near the doubly-magic $\Nuc{Sn}{132}{50}{82}$ isotope.
Further theoretical studies have revealed that the evolution of
deformed shell effect in the fission process is essential in
determining the fragment distribution, and the shell effect associated
with spatially localized prefragments should be present.  However, the
standard quantum mechanical mean-field approaches cannot extract such
prefragment shell effect out of that in the total system.  The parity
splitting of levels in the two-center shell model potential is
investigated as the indication of fragment shell
effect\cite{Mosel71a,Mosel71b}, but it is limited to symmetric shapes.
Shell effects of the independent fragments are discussed in some
recent works\cite{Scamps18,Scamps19}, but it is not trivial to clarify
how they reflect the effect of the prefragment embedded in the total
system.  We should say that the physical mechanism of the asymmetric
fission has not been sufficiently clarified.  Since the asymmetry
itself can be reproduced in mean-field calculations, its origin must
be explained within the mean-field theory.  However, it is hard to
define the shell effect associated with each of the prefragments
because most of the single-particle wave functions are delocalized in
the mean-field potential.

It was pointed out by Strutinsky et al. that the semiclassical
periodic-orbit theory (POT) \cite{Gutz,BB3} would be able to explain
the origin of such prefragment shell effect\cite{StrMag76}.  In the
POT, single-particle level density is represented as the sum over the
contributions of classical periodic orbits (POs).  If a neck is formed
upon the elongated potential, a set of POs appear which are confined
in each of the prefragments, and their contributions to the level
density can be regarded as the prefragment shell effect.  However,
such kind of analysis hasn't been carried out in deformed potential
models for the fission processes.  In this work, we apply the POT to a
simple deformed cavity potential model and discuss the effect of
prefragment shell effect to elucidate the underlying mechanism of
asymmetric fission.

\section{Asymmetric fission}
\label{sec:exp}

\begin{figure}
\includegraphics[width=\linewidth]{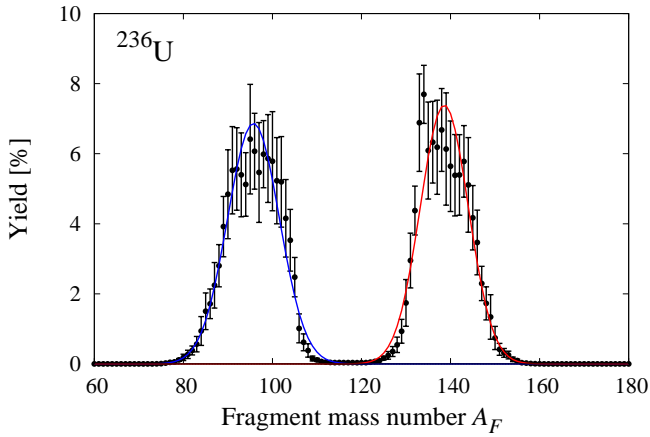}
\caption{\label{fig:fmdU236}
Fragment mass distribution in the neutron-induced fission of
\nuc{U}{236}.  Dots with error bars represent the experimental
data taken from ENDF nuclear database\cite{ENDF}.  Solid curves show
the Gaussians fitted to each of the two peaks corresponding to
the heavier and lighter fragments.}
\end{figure}

Figure~\ref{fig:fmdU236} shows the distribution of fragment mass
yields for the neutron-induced fission of \nuc{U}{236}.  It
consists of two peaks with the heavier component around $A=140$ and
the lighter component around $A=96$.  In the figure, each of the peaks
is fitted by the Gaussian.
\begin{figure}
\includegraphics[width=\linewidth]{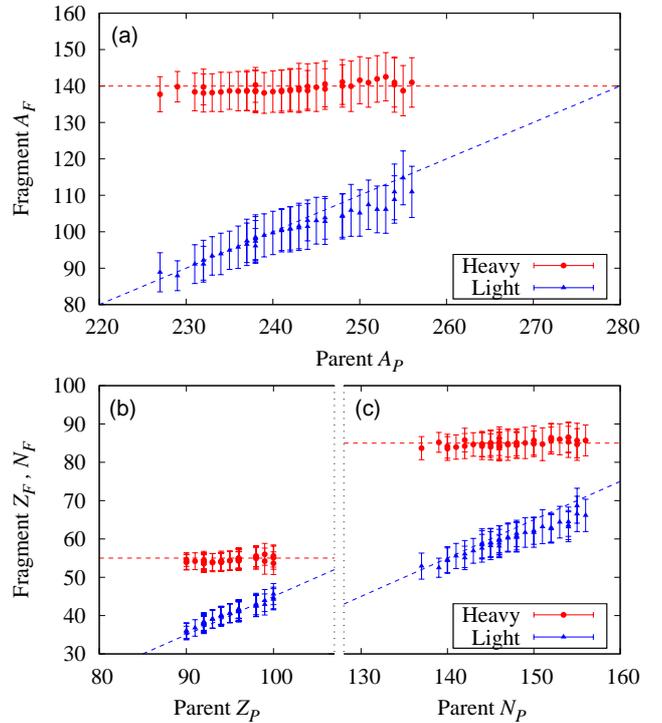}
\caption{\label{fig:fmd}
Fragment mass-number distribution in the spontaneous and
neutron-induced fissions of actinide nuclei is shown in the panel (a).
Dots show the center of the Gaussians that fit the heavier and lighter
components of the experimental fragment mass-number distributions (see
Fig.~\ref{fig:fmdU236}), and the standard deviations around them are
indicated by the vertical bars.  The same plots as the panel (a) but
for proton and neutron numbers are shown in the panels (b) and (c),
respectively.  Broken lines represent $A_F=140$ and $A_F=A_P-140$ in
the panel (a), $Z_F=55$ and $Z_F=Z_P-55$ in the panel (b), and
$N_F=85$ and $N_F=N_P-85$ in the panel (c).}
\end{figure}
This two-peak structure in the fragment-mass distribution is common
among all other actinide nuclei.  The positions of the centers of
those peaks and the standard deviations around them are summarized in
Figure~\ref{fig:fmd}.  The outstanding feature is that the peak of the
heavier fragment is always found around $A\sim 140$ ($Z\sim 55$ and
$N\sim 85$) independently of the parent species.  Since these numbers
are close to the magic numbers $Z=50$ and $N=82$, the energy gain due
to the shell effect of the fragment has been considered as the primary
driving force behind the above asymmetric fissions.

For lighter nuclei, the fissions had been expected to occur in more
symmetric manners because such strong shell effect that advantages the
fragment mass asymmetry seems to be absent.  However, substantially
asymmetric fragment mass distributions were observed in the fissions
of some neutron-deficient mercury isotopes.  In the relatively recent
experiment\cite{Andr2010}, the fission of \nuc{Hg}{180} has turned out
to be asymmetric although the symmetric fission product
\Nuc{Zr}{90}{40}{50} with neutron number at magic $N=50$ and
proton number at submagic $Z=40$ is very stable.

The origin of the above asymmetry, what they call a new type of asymmetric
fission, has been theoretically studied in several approaches: see
\cite{Ichikawa2012,Warda2012} for instance.  These works have pointed out
the significance of finding the optimum fission path on the potential
energy surface which runs through the normally deformed ground state
and the elongated saddle points.  In the case of \nuc{Hg}{180}, one
finds a deep valley along the line of the symmetric shapes at large
elongation in the potential energy surface due to the low energy of
the symmetric fission products \nuc{Zr}{90}.  But it is inaccessible
from the fission path consisting of a sequence of minima and saddles
from the normal deformed minima because they are separated from each
other by a high potential ridge (see, e.g., Fig.~7 of
Ref.~\cite{Ichikawa2012}).  It tells us that the energies of the final
states alone are not sufficient to understand the asymmetric fission.
It is also important to consider the potential landscape in the shape
parameter and find the energetically favored fission path along which
the shape of the system likely to evolve towards the scission.

In Ref.~\cite{Warda2012}, it is found that a prefragment whose density
distribution is quite similar to that of isolated \nuc{Zr}{90} comes
up in the elongated parent nucleus \nuc{Hg}{180}.  With such a
configuration, the other prefragment necessarily becomes lighter since
there must be some nucleons in the neck part between the two
prefragments.  From this observation, they concluded that the shell
effect associated with the prefragment corresponding to \nuc{Zr}{90}
plays a role to make the fission asymmetric rather than symmetric,
contrary to the first expectation.  More recently, nucleon
localization functions\cite{NLF} were investigated in the microscopic
calculations for fission deformation processes\cite{Naz1,Naz2}, which
clearly indicate the formation of prefragments similar to relatively
stable isolated nuclei.  For fissions of superheavy nuclei, it is
predicted that the shell structure of doubly magic \nuc{Pb}{208}, as
well as that of \nuc{Sn}{132}, plays an important
role\cite{Ichikawa2005,Warda2018,Matheson2019}
(see also \cite{Sharma1,Sharma3}).

The above numerical outcomes indicate the significance of shell effect
associated with prefragments formed in the elongated nuclear body.  In
those works, realistic models are used which take into account various
effects such as Coulomb force, pairing correlations, and realistic
nucleon distributions.  Those effects are all important to reproduce
the individual experimental data.  However, to answer the fundamental
question what the essential mechanism for the asymmetric fission is,
it may be useful to study a simplified model that captures the essence
of the relation between the shell evolution and the shape change
during the fission process.  In the following part, we shall use
extremely simplified mean-field potential model to focus our attention
on the role of the shape evolution in the fission deformation
processes.  The prefragment shell effect is considered by the POT in
line with Strutinsky's view\cite{StrMag76}.

\section{Three-Quadratic-Surfaces parametrization}
\label{sec:tqs}

For describing the fission deformation processes, several types of
shape parametrization have been proposed.  Two-center shell model
potential, consisting of two oscillators centered at two different
points and the neck part smoothly connecting them, have been utilized
in several static and dynamical
calculations\cite{Mosel71a,Mosel71b,Aritomo2014}.  It includes the
five essential parameters to describe the shape of the potential:
elongation, fragment mass asymmetry, neck radius, and quadrupole
deformations of the two prefragments.  Although it is important to
fully consider those five shape degrees of freedom, parameter sets
with reduced numbers have also been used for simpler analyses.  The
($c$-$h$-$\alpha$) model with three parameters controlling the
elongation, neck shape and asymmetry, was employed in the review
article \cite{FunnyHills} on the application of the shell correction
method to the fission problem.  Semiclassical analysis was made in the
cavity model with the same shape parametrization\cite{Brack97}, and
the role of POs in generating fission path leading to the asymmetric
shape was discussed.

\begin{figure}
\includegraphics[width=\linewidth]{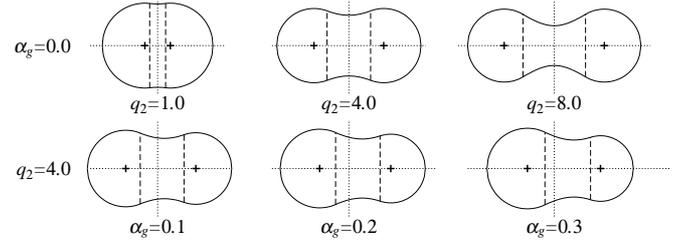} %{shape1_TQS.eps}
\caption{\label{fig:shape_tqs}
Shapes of the 3QS surface with several values of the elongation
parameter $q_2$ and prefragment mass asymmetry $\alpha_g$.
Prefragments deformation parameters are put to $\epsilon_j=0$
(spherical) and the neck parameter is fixed at $\sigma_2=-0.6$.  The
vertical broken lines represent the joints between adjacent quadratic
surfaces.  Dotted lines indicate the symmetry axis and the position of
the center of mass.}
\end{figure}

Since our aim is to discuss the prefragment shell effect, the
three-quadratic-surfaces (3QS) parametrization proposed by
R.~Nix\cite{Nix69} is convenient with which one can easily control the
shapes of the prefragments\cite{MSI2009}.  In the 3QS, the surface of
the axially symmetric potential $\rho=\rho_s(z)$ is divided into three
regions along the axis of symmetry direction, and each of them is
expressed as a quadratic surface,
\begin{equation}
\rho_s^2(z)=\left\{\begin{array}{l@{\quad}l}
a_1^2-\frac{a_1^2}{c_1^2}(z-l_1)^2, & (l_1-c_1\leq z\leq z_1) \\
a_3^2-\frac{a_3^2}{c_3^2}(z-l_3)^2, & (z_1< z< z_2) \\
a_2^2-\frac{a_2^2}{c_2^2}(z-l_2)^2. & (z_2\leq z\leq l_2+c_2)
                      \end{array}\right.
\end{equation}
These three parts are smoothly connected at the joints $z=z_1$ and
$z_2$.  The established surface is described by the five independent
shape parameters $\{q_2,\alpha_g,\sigma_2,\epsilon_1,\epsilon_2\}$
under the center-of-mass and volume-conservation conditions.  $q_2$ is
the dimensionless elongation parameter proportional to the quadrupole
moment $Q_2$, defined as\cite{CS1963}
\begin{equation}
Q_2=\int \rho_c(\br)(2z^2-x^2-y^2)dV
=\frac{ZeR_0^2}{4\pi/3}q_2,
\end{equation}
where the charge density $\rho_c(\br)$ is assumed to be uniform inside
the surface, and $R_0$ is the nuclear radius in the spherical limit.
$\alpha_g$ is the prefragment mass asymmetry
\begin{equation}
\alpha_g=\frac{M_1-M_2}{M_1+M_2}
=\frac{a_1^2c_1-a_2^2c_2}{a_1^2c_1+a_2^2c_2},
\end{equation}
where the mass $M_j$ of the $j$th prefragment ($j=1,2$) is calculated
assuming a uniform-density spheroidal body.  $\sigma_2$ is the
curvature of the middle surface
\begin{equation}
\sigma_2=\frac{a_3^2}{c_3^2},
\end{equation}
which takes negative values ($c_3^2<0$) when the neck is formed and
the nuclear surface turns a dumbbell shape.  $\epsilon_j$ $(j=1,2)$ is
the spheroidal deformation parameters of the $j$th prefragment,
\begin{equation}
\epsilon_j=\frac{3(c_j-a_j)}{2a_j+c_j}.
\end{equation}
In the present work, we shall fix the shapes of the prefragments to be
spherical $(\epsilon_j=0)$.  We also fix the neck parameter to
$\sigma_2=-0.6$, which is close to its values for some actinide nuclei
along the fission paths obtained in the realistic
macroscopic-microscopic calculations\cite{Ichikawa2012}.  Then, we
consider the deformed shell structure against the elongation and
fragment mass asymmetry.  Shapes at several values of
$\{q_2,\alpha_g\}$ are displayed in Fig.~\ref{fig:shape_tqs}.
Consideration of the roles of the prefragment deformations is left for
future studies.

Using this parametrization, let us first examine the deformation
energy in the liquid-drop model (LDM).  The LDM deformation energy
consists of surface and Coulomb parts
\begin{equation}
\Delta E_{\rm LDM}(q)=b_S(q) A^{2/3}+b_C(q) \frac{Z^2}{A^{1/3}},
\end{equation}
where the coefficients $b_S(q)$ and $b_C(q)$ are dependent on the
deformation $q=\{q_2,\alpha_g\}$.  

\begin{figure}
\includegraphics[width=\linewidth]{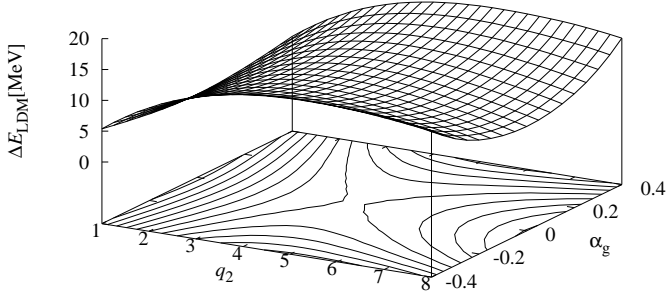} %{ldm1_U236.eps}
\caption{\label{fig:ldm1}
LDM deformation energy for \nuc{U}{236} as function of the elongation
parameter $q_2$ and the asymmetry parameter $\alpha_g$.  Contour
curves are shown on the bottom plane.}
\end{figure}

Figure~\ref{fig:ldm1} shows the LDM deformation energies for the
nucleus \nuc{U}{236} as functions of the elongation parameter $q_2$
and the asymmetry parameter $\alpha_g$, with the prefragment
deformations and the neck parameter fixed at $\epsilon_1=\epsilon_2=0$
and $\sigma_2=-0.6$.  One sees that the asymmetric configuration
$(\alpha_g>0)$ is disfavored through the fission deformation
processes.  Thus, the origin of the asymmetric fission cannot be found
in the classical LDM energy.

\section{Shell structures in the fission processes
and the periodic-orbit theory}
\label{sec:shell}

In this section, we first give a brief introduction to the general
aspects of semiclassical periodic orbit theory.  The advantage of the
semiclassical theory in considering the prefragment shell effect is
emphasized.  Then, we apply the theory to the 3QS cavity model and
investigate the shell structures in the fission deformation processes.

\subsection{Semiclassical theory of shell structures}
\label{sec:trace}

The single-particle level density for the mean-field Hamiltonian
$\hat{h}$ is given by
\begin{equation}
g(e)=\sum_n\delta(e-e_n)=\int dt e^{iet/\hbar}\int d\br
K(\br,\br;t),
\end{equation}
where $e_n$ is the $n$th energy eigenvalue of $\hat{h}$ from the bottom,
and $K(\br',\br;t)=\<\br'|e^{-it\hat{h}/\hbar}|\br\>$ is the transition
amplitude which can be expressed in
the path integral representation.  Semiclassical evaluation of the
path integral using the stationary-phase method extracts
contributions of classical POs, and one obtains the so-called trace
formula which expresses the quantum level density in terms of the
classical POs as
\begin{equation}
g(e)\simeq \bar{g}(e)+\sum_\beta
A_\beta(e)\cos\left(\frac{1}{\hbar}S_\beta(e)
-\frac{\pi}{2}\mu_\beta\right).
\label{eq:trace_formula}
\end{equation}
In the right-hand side, $\bar{g}(e)$ represents the average part of
the level density which is generally a moderate and monotonous
function of energy, and the second term represents the oscillating
part.  The sum is taken over all the POs in the classical counterpart
of the system, where $\beta$ specifies the orbit.
$S_\beta(e)=\oint_\beta\bp\cdot d\br$ is the action integral along the
orbit $\beta$, $\mu_\beta$ is the Maslov index related to the
geometric property of the orbit, and the amplitude $A_\beta$ is
determined by the degeneracy, stability and period of the orbit.

In a cavity potential, the classical particle moves rectilinearly and
is reflected ideally at the boundary, and one has the same set of POs
independent of energy.  Since the modulus of momentum $p=\hbar k$ is
constant, action integral along the orbit is simply given by
\begin{equation}
S_\beta(k)=\hbar kL_\beta,
\end{equation}
where $L_\beta$ is the geometric length of the orbit.  In this case,
it is more useful to rewrite the trace formula
(\ref{eq:trace_formula}) in terms of the wave number variable $k$,
instead of energy $e$:
\begin{align}
g(k)&=g(e)\frac{de}{dk} \NN
&\simeq \bar{g}(k)+\sum_\beta A_\beta(k)
\cos\left(kL_\beta-\frac{\pi}{2}\mu_\beta\right).
\label{eq:trace_k}
\end{align}
The contribution of each orbit gives a regularly oscillating function
of $k$, and the period of the oscillation $\delta k$ is inversely
proportional to the length of the orbit
\begin{equation}
\delta k=\frac{2\pi}{L_\beta}.
\end{equation}
Accordingly, the shorter orbits are responsible for the gross shell
effect with large $\delta k$.  Generally a complicated structure in
the level density fluctuation is built up with the superposition of the
contributions of various orbits having different lengths.
\begin{figure}
\includegraphics[width=\linewidth]{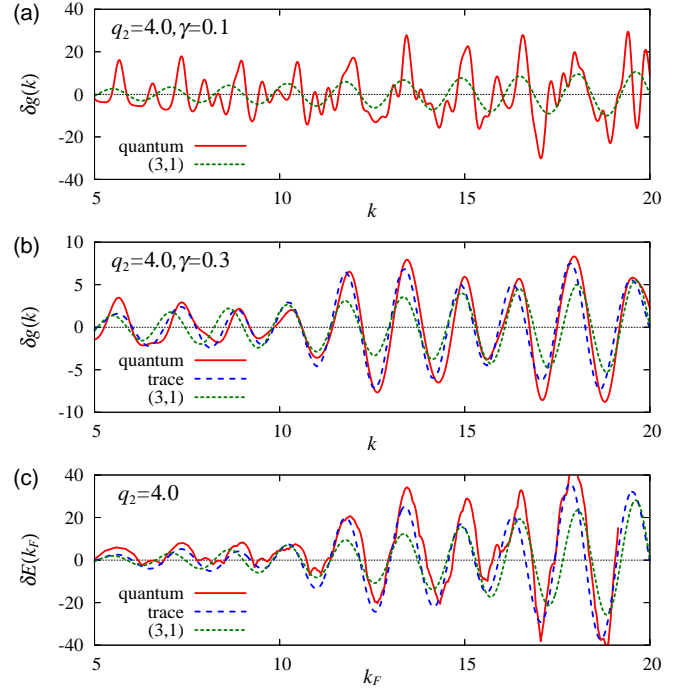} %{sld_q40.eps}
\caption{\label{fig:sld}
Comparison between quantum and semiclassical calculations for the
shell structures in the symmetric 3QS cavity model with the elongation
parameter $q_2=4.0$.  In the panel (a) and (b), oscillating parts of
the single-particle level densities (\ref{eq:sld}) with the averaging
parameter $\gamma=0.1$ and $0.3$ are shown as functions of the wave
number $k$.  In the panel (c), shell energy is plotted as a function
of the Fermi wave number $k_F$.  The solid (red) curves represent the
quantum results, and the long-dashed (blue) curves represent the
results of trace formula (\ref{eq:trace_k}) and (\ref{eq:sce}).  The
short-dashed (green) curves show the contribution of the triangle
orbit family (3,1) in the prefragments (discussion will be conducted
in Sec.~\ref{sec:pot}).}
\end{figure}

In Fig.~\ref{fig:sld}, we show the oscillating part of the level
density coarse-grained with the averaging width $\gamma$,
\begin{equation}
\delta g_\gamma(k)=\int dk' \{g(k')-\bar{g}(k')\}
 \exp\left[-\frac12\left\{\frac{(k'-k)R_0}{\gamma}\right\}^2\right]
\label{eq:sld}
\end{equation}
for the symmetric 3QS shape with the elongation parameter $q_2=4.0$.
With the averaging width $\gamma$, contributions of long POs having
the lengths $L_\beta\gtrsim R_0/\gamma$ are integrated out and only
the shorter POs prevail.  With $\gamma=0.1$, many orbits up to $L\sim
10R_0$ contribute and one sees a complicated fine structure.  With
$\gamma=0.3$, one finds a simple oscillating pattern governed by only
a few shortest orbits.  In the panel (b) of Fig.~\ref{fig:sld}, one
sees that the gross structures in quantum level densities are nicely
reproduced by the semiclassical trace formula taking account of only
five shortest PO families confined in the prefragments.

Shell energy is directly related to the oscillating part of the level
density as\cite{FunnyHills}
\begin{equation}
\delta E(N)=\int^{e_F}(e-e_F)\delta g(e)de,
\end{equation}
with the Fermi energy $e_F$ satisfying
\begin{equation}
\int^{e_F}g(e)de=N.
\end{equation}
$\delta E$ depends essentially on the gross shell structures of
$\delta g$ since the fine structures are mostly integrated out.  By
inserting the PO sum (\ref{eq:trace_formula}) into $\delta g(e)$, one
obtains the semiclassical formula for shell
energy\cite{StrMag76,BrackText}
\begin{equation}
\delta E(N)=\sum_\beta\frac{\hbar^2}{T_\beta^2}A_\beta(e_F)
\cos\left(k_FL_\beta-\tfrac{\pi}{2}\mu_\beta\right),
\label{eq:sce}
\end{equation}
with the Fermi wave number $k_F=\sqrt{2Me_F}/\hbar$.  In
Eq.~(\ref{eq:sce}), the contribution of each PO has an additional
factor proportional to $T_\beta^{-2}$, which plays a role to suppress
the contributions of longer POs.  In the panel (c) of
Fig.~\ref{fig:sld}, shell energy of the 3QS cavity system with the
same shape as that used in the two upper panels is plotted as a
function of the Fermi wave number.  One sees that the oscillating
pattern is nicely reproduced simply by the contributions of some
shortest POs.

In general, shell structures are known to be very sensitive to the
shape of the potential.  In semiclassical point of view, it can be
explained by the sensitivity of the stability of POs to the potential
shape, as well as the changes of the orbit lengths which lead to the
different kinds of interference effects.

\subsection{Prefragment shell effect --- relation to classical
periodic orbits}
\label{sec:pot}

\begin{figure}
\includegraphics[width=\linewidth]{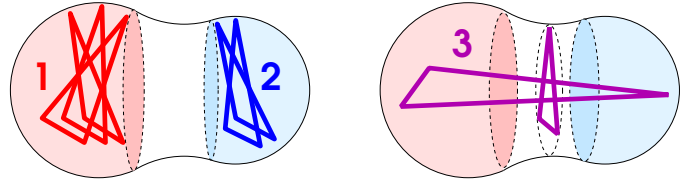}
\caption{\label{fig:po_frag}
Classification of classical POs in the fissioning
cavity potential model.}
\end{figure}

As stated above, one can extract the prefragment contribution out of
the total shell energy using the POT.  Periodic orbits in our model
can be classified into the following three groups:
\begin{enumerate}
\item orbits confined in the 1st prefragment
\item orbits confined in the 2nd prefragment
\item orbits staying in the middle surface or those traveling
between two prefragments
\end{enumerate}
as illustrated in Fig.~\ref{fig:po_frag}.
An unambiguous definition of the prefragment shell effect can be given
by the contributions of orbits confined in the corresponding
prefragment.  According to the above classification of POs, we
decompose the shell energy into three parts as
\begin{equation}
\delta E(N)={\delta E_1(N)}+
            {\delta E_2(N)}+
            {\delta E_3(N)}.
\end{equation}
The orbits included in the third category generally have less
contributions to the shell energy because of the small degeneracy
compared to the prefragment orbits.  Thus, the prefragment shell
effect dominates the total shell effect with developing neck
configuration.  In this way, shape stabilities of the prefragments are
expected to play a crucial role in the deformed shell effect in the
fission process.

To estimate the prefragment shell effect, one has to calculate the
classical POs and their characteristics such as periods, degeneracies
and stabilities.  All the classical periodic orbits in the spherical
cavity potential are obtained analytically.  They are specified by the
two indices $(p,t)$: $p$ counts the number of reflections on the
surface, and $t$ the number of rotations around the center of the
sphere.  Regular polygon orbits $(p>2t)$ such as triangle (3,1) and
square (4,1) orbits form three-parameter families generated by the
three-dimensional rotation, while the diameter orbits $(p=2t)$ form
two-parameter families\cite{BB3}.  In the 3QS cavity potential under
consideration, one has the same diameter and polygon orbits confined
in prefragments which are truncated spheres.  By considering the
restricted ranges of rotation angles for those orbit families in the
prefragments, the reduction factor $f_p$ of the amplitude relative to
that for the family in non-truncated spherical cavity can be obtained.
The principal part of the contribution of the PO family is given by
the amplitude
\begin{equation}
A_{pt}^{\rm (pr)}=f_p A_{pt}^{\rm (sph)}.
\end{equation}
However, the above contribution is insufficient to reproduce the
quantum results, and one needs to take into account the end-point
corrections to the contribution of the truncated family.  By extending
the Balian-Bloch trace formula\cite{BB3}, we have derived the formula
for the contribution of such truncated family \cite{Arita2018,AIM2018}
in the form
\begin{align}
\delta g_{pt}(k)&=\sum_D A_{pt}^{(D)}(k)\cos(kL_{pt}
 -\tfrac{\pi}{2}\mu^{(D)}_{pt}) \NN
&=\Re\Biggl[\Biggl(\sum_D A_{pt}^{(D)}
 e^{-i\frac{\pi}{2}\mu_{pt}^{(D)}}\Biggr)e^{ikL_{pt}}\Biggr] \NN
&\equiv A_{pt}(k)\cos(kL_{pt}-\tfrac{\pi}{2}\mu_{pt}), \label{eq:trace_trpo}
\end{align}
where the sum in the first line is taken over the principal part (with
degeneracy $D=D_{\rm max}$) and several orders of the end-point
corrections (with $D<D_{\rm max}$).  In the next subsection, we will
compare the quantum results with our semiclassical formula.

\subsection{Quantum-Classical correspondence in Fourier transformation}
\label{sec:fourier}

In the cavity model, one can easily extract information on the
contributions of classical POs by Fourier transformation of quantum
level density.  Practically, quantum spectra is available up to a
finite maximum value, and we truncate the high-energy part of the
spectrum with the Gaussian and consider the following Fourier
transform:
\begin{equation}
F(L)=\sqrt{\frac{2}{\pi}}\frac{1}{k_c}
\int dk\,g(k) e^{ikL} e^{-\frac12(k/k_c)^2}.
\label{eq:ftl}
\end{equation}
Inserting the trace formula (\ref{eq:trace_k}), one obtains
\begin{align}
F^{sc}(L)
&=F_0(L)+\sum_\beta A_\beta
 e^{-i\pi\mu_\beta/2} \NN
&\quad\times \exp\left[-\frac12\{k_c(L-L_\beta)\}^2\right].
\label{eq:ftl_sc}
\end{align}
Using this relation, one can extract information on the contributions
of classical POs out of the quantum spectrum.  The modulus of the
Fourier transform exhibits successive peaks centered at the lengths of
the classical POs $L=L_\beta$, and the amplitude $A_\beta$ of a
certain orbit $\beta$ is available from the height of the peak by
\begin{equation}
A_\beta \approx |F(L_\beta)| \label{eq:peak}
\end{equation} 
if there is no other peak in the vicinity.  Eq.~(\ref{eq:ftl_sc}) is
derived by ignoring the $k$ dependence of the amplitude for
simplicity.  The expression taking account of the correct $k$
dependence is given in \cite{Arita2018,AIM2018}, which just replaces
the gaussian with another similar function.

\begin{figure}
\centering
\includegraphics[width=\linewidth]{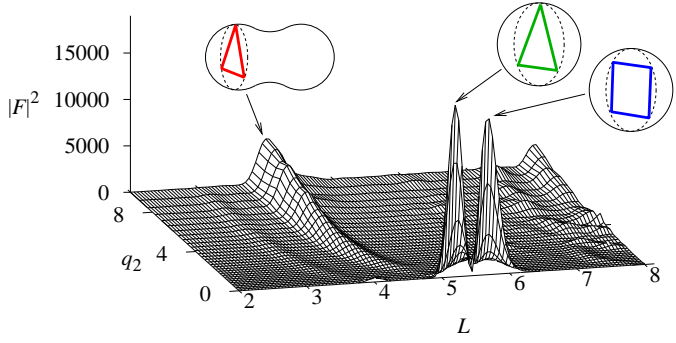} %{fourier_N06D100n.eps}
\caption{\label{fig:fourier}
Squared moduli of the Fourier transform of quantum level density,
$|F(L;q_2)|^2$, plotted as a function of the length parameter $L$
(in unit of $R_0$) and the elongation parameter $q_2$.  The
classical POs associated with the peaks are
indicated by the inserted pictures.}
\end{figure}

Figure~\ref{fig:fourier} shows the Fourier transform of the quantum
mechanical level density for the symmetric 3QS cavity.  Squared
modulus of the Fourier amplitude, $|F(L;q_2)|^2$, calculated for
varying $q_2$ is plotted on the $(L,~q_2)$ plane.  At the spherical
shape, $q_2=0$, one sees two pronounced peaks corresponding to the
triangle and square PO families having the lengths
$L_{31}=3\sqrt{3}\simeq 5.20$ and $L_{41}=4\sqrt{2}\simeq 5.66$,
respectively, in units of $R_0$.  The peak corresponding to the
diameter orbit at $L=L_{21}=4$ is much smaller because of the small
degeneracy.  With increasing $q_2$, the above two peaks rapidly
decrease, and instead, a peak corresponding to the prefragment
triangle orbit grows up and makes a significant contribution at large
$q_2$.

\begin{figure}
\centering
\includegraphics[width=\linewidth]{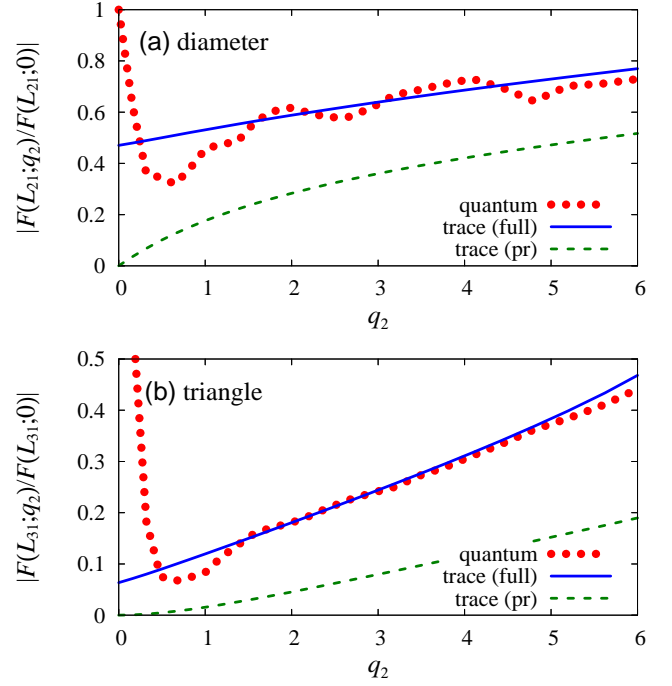} %{ftl_obtn.eps}
\caption{\label{fig:ftl_obt}
Quantum Fourier amplitudes at the lengths of the PO (dotted curves)
compared with the semiclassical trace formula for (a) the diameter and
(b) the triangle orbit families confined in the spherical
prefragments.  Their values relative to those for non-truncated
spherical cavity are plotted as functions of the elongation parameter
$q_2$.  The solid curves represent the results of the trace formula
where the end-point corrections are fully taken into account.  The
dashed curves represent the principal parts of the semiclassical
amplitudes alone, namely, the reduction factors $f_2$ and $f_3$ (see
Sec.~\ref{sec:pot}) in the panels (a) and (b), respectively.}
\end{figure}

In Fig.~\ref{fig:ftl_obt}, the Fourier peak of the quantum level
density at the lengths of the prefragment diameter (2,1) and triangle
(3,1) orbits are compared with the semiclassical amplitudes derived in
\cite{Arita2018}, according to Eq.~(\ref{eq:peak}).  As we discussed
in Sec.~\ref{sec:pot}, one has the same families of the diameter and
regular polygon orbits in the prefragments as those in the
non-truncated spherical cavity, but with the restricted ranges of the
parameters.  The dotted curves in Fig.~\ref{fig:ftl_obt} represent the
moduli of the Fourier transform (\ref{eq:ftl}) at the length of the
diameter and triangle POs, divided by their values at the spherical
shape $q_2=0$.  The corresponding semiclassical results shown by the
solid lines are the amplitudes $A_{pt}$ in Eq.~(\ref{eq:trace_trpo}),
including principal parts and all the end-point corrections, divided
by those for non-truncated spherical cavity.  The dashed curves show
the principal parts, namely, the reduction factor $f_p$ of the
truncated family $(p,t)$, which considerably underestimate the quantum
results.  By taking into account the end-point corrections, quantum
results are nicely reproduced both for the diameter and triangle POs.

\subsection{Semiclassical analysis of the prefragment shell effect}
\label{sec:result}

In the following part, we investigate the prefragment shell effect in
the 3QS cavity model using the trace formula (\ref{eq:sce}) for shell
energy.

\begin{figure}
\includegraphics[width=\linewidth]{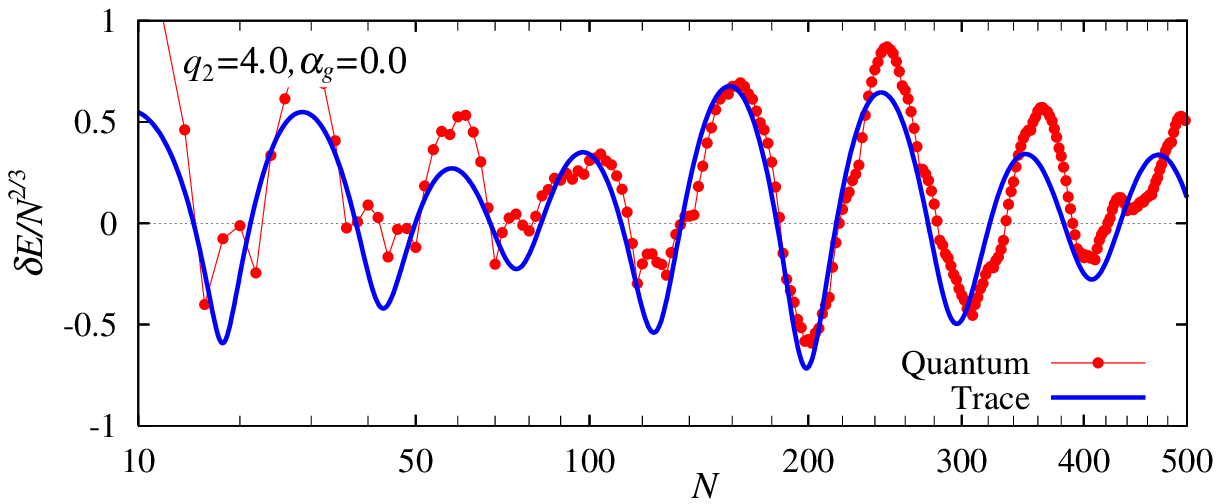} %{sce_sig1n.eps}
\caption{\label{fig:sce_s1}
Shell energy $\delta E(N)$ divided by $N^{3/2}$ as a function of
particle number $N$ for the symmetric 3QS cavity model with the
elongation parameter $q_2=4.0$.  The thin solid line with dots (in
red) represents the quantum result and the thick solid curve (in blue)
represents the semiclassical trace formula taking some shortest
prefragment POs into account.}
\medskip

\includegraphics[width=\linewidth]{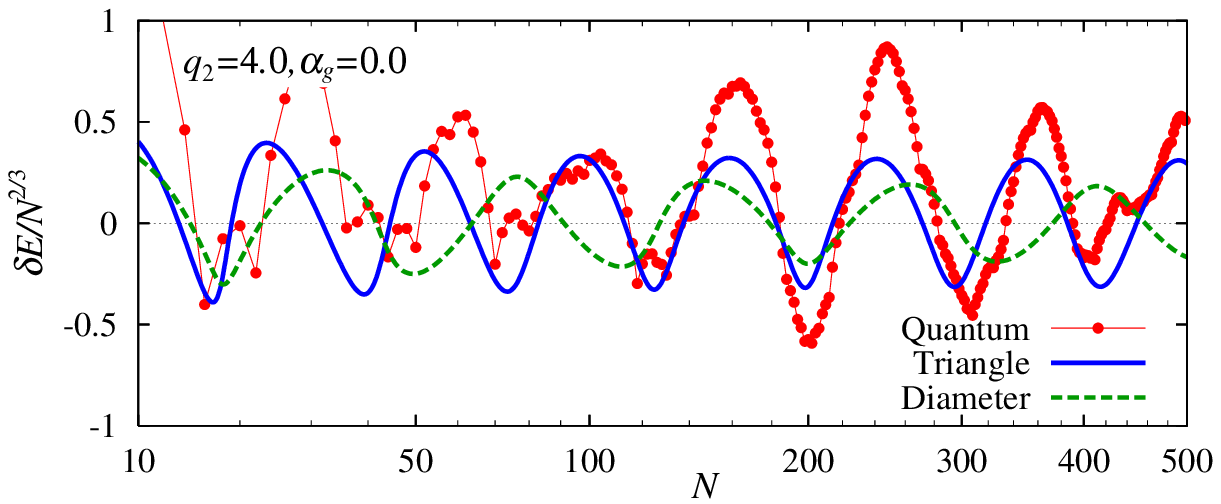} %{sce_sig2n.eps}
\caption{\label{fig:sce_s2}
Decomposition of the shell energy into the contribution of individual
PO for the symmetric 3QS cavity model with the same shapes as in
Fig.~\ref{fig:sce_s1}.  Thin solid line with dots represents the
quantum results equivalent to the one in Fig.~\ref{fig:sce_s1}, solid
and dashed curves represent the contributions of triangle and diameter
orbits, respectively.}
\end{figure}

Figure~\ref{fig:sce_s1} shows the shell energy $\delta E(N)$ for the
symmetric 3QS cavity model, where POs confined in one of the
prefragments and those in the other are equivalent, and they make
constructive contributions.  Quantum results are nicely reproduced by
the semiclassical trace formula taking into account the five shortest
prefragment POs.  One finds a modulation in the gross shell structure.
This modulation is caused by the interference between POs with
different lengths.  In Fig.~\ref{fig:sce_s2}, contributions of the
diameter and triangle orbits are shown.  The diameter orbit has small
amplitude in the Fourier analysis (see Fig.~\ref{fig:fourier}) but it
has significant contribution to the shell energy due to the shortness
[see Eq.~(\ref{eq:sce})].  Especially deep minima around $N=200$ for
$q_2=4.0$ are caused by the constructive contributions of those two
orbits.  For particle numbers where those contributions are
destructive, shell effect becomes relatively weak.

\begin{figure}
\includegraphics[width=\linewidth]{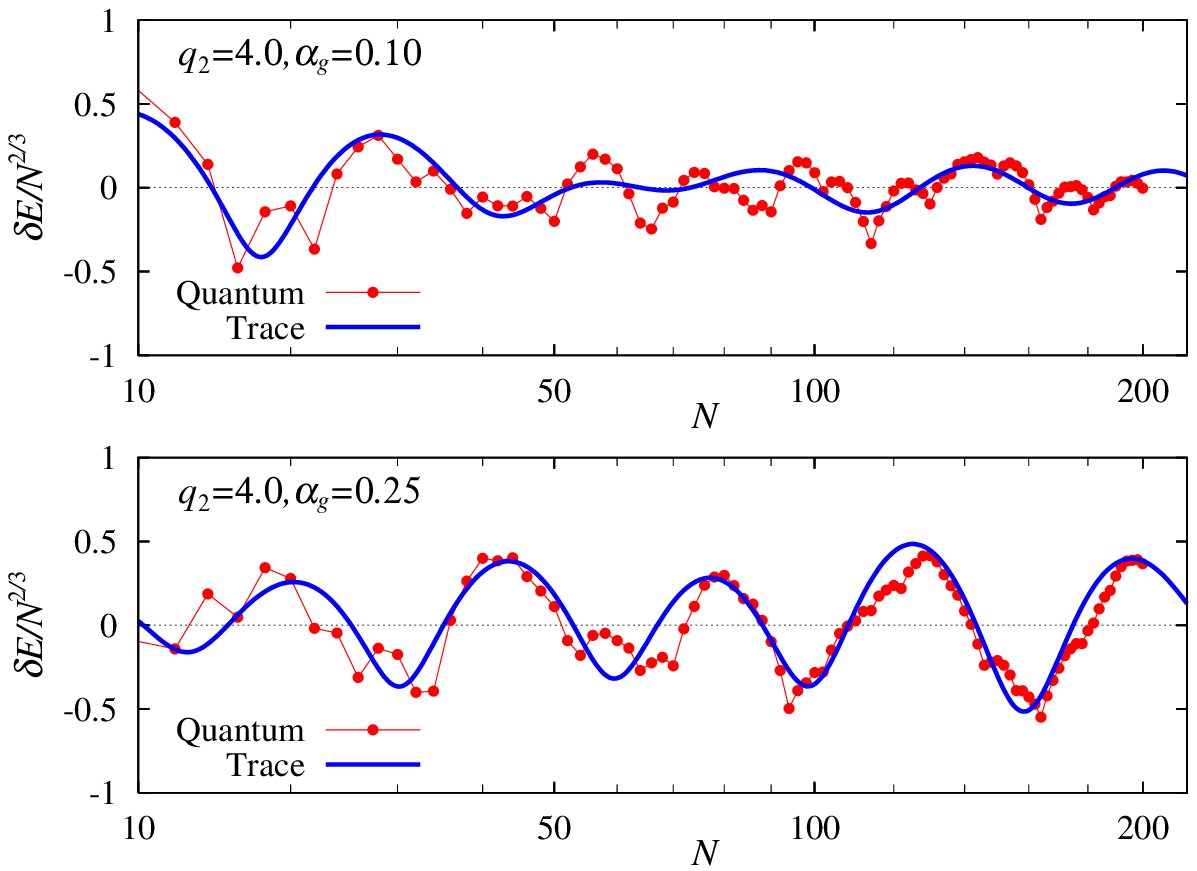}
\caption{\label{fig:sce_a1}
Shell energy of the 3QS cavity model with the elongation parameter
$q_2=4.0$ and asymmetry parameter $\alpha_g=0.1$ and $0.25$.
The thin line with dots represents quantum result, and the thick solid
curve represents the result of semiclassical trace formula taking account of
some shortest prefragment POs.}
\medskip

\includegraphics[width=\linewidth]{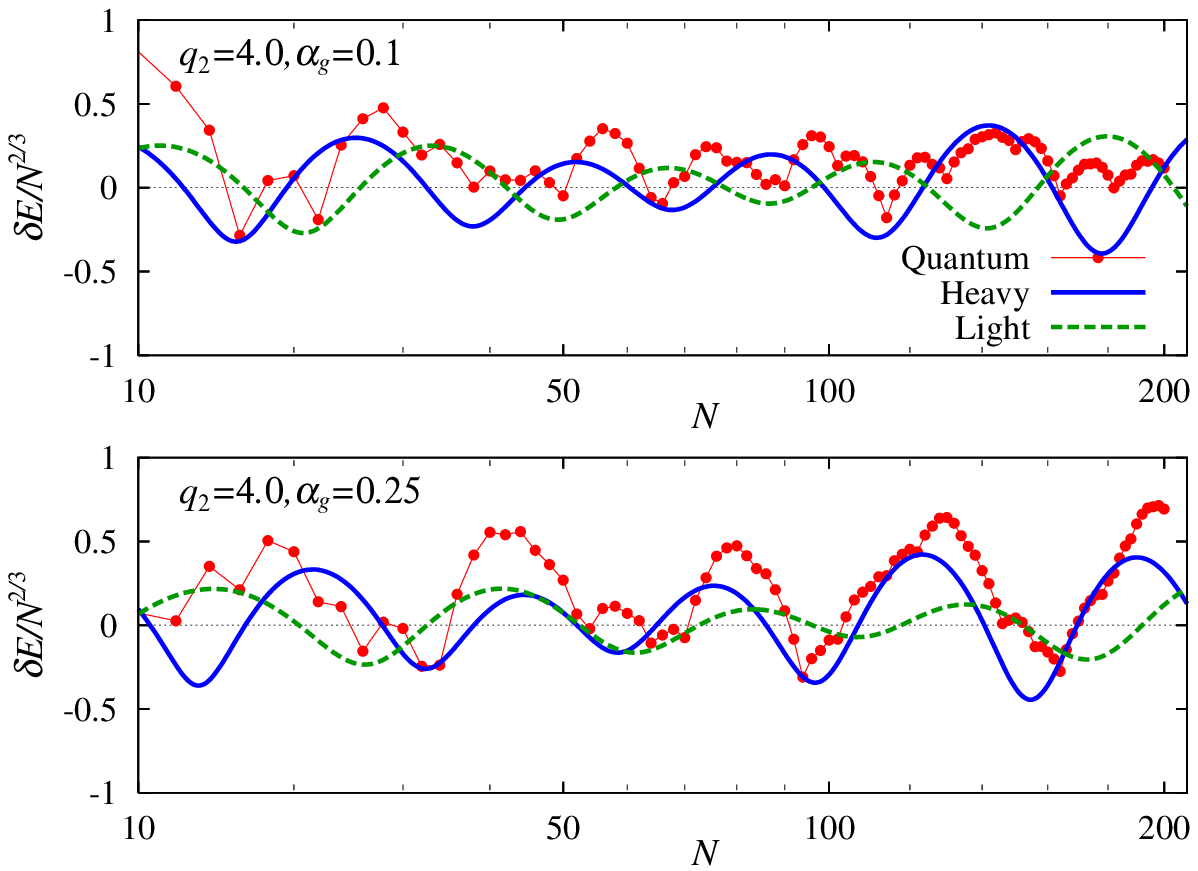}
\caption{\label{fig:sce_a2}
Decomposition of the shell energy into the contributions of POs in
each of the prefragments for the asymmetric 3QS cavity model with the
same shapes as those in Fig.~\ref{fig:sce_a1}.  The thin line with
dots represents the quantum results equivalent to the one in
Fig.~\ref{fig:sce_a1}.  Solid and dashed curves represent the
contributions of POs in heavy and light prefragments, respectively.}
\end{figure}

The results for asymmetric shapes ($\alpha_g>0$) with $q_2=4.0$ are
shown in Figs.~\ref{fig:sce_a1} and \ref{fig:sce_a2}.
Quantum shell energy is nicely reproduced by the trace formula in
Fig.~\ref{fig:sce_a1}, except for some fine structures.  In
Fig.~\ref{fig:sce_a2}, contributions of heavy and light prefragments
are shown.  Due to the asymmetry, the orbits of the same type but in
the different prefragments have different lengths, and one finds an
interference between them.  For $\alpha_g=0.1$, contributions of two
prefragments are out of phase in the plotted particle number region
and the shell effects are relatively weak.  For $\alpha_g=0.25$, those
contributions turn more constructive and one finds larger shell
effects.

\subsection{Effect of prefragment magics on asymmetric fission}
\label{sec:magic}

Since each spherical prefragment has the same set of PO families as in
the non-truncated spherical cavity, one can expect the possibility of
expressing the shell energy of the 3QS cavity in terms of the
spherical one.  Let us define the factor $w_\beta^{(j)}$ by
\begin{equation}
A_\beta^{(j)}(e)\simeq w_\beta^{(j)}A_\beta^{\rm(sph)}(e;R_j).
\end{equation}
It represents the value of the amplitude $A_\beta^{(j)}$ for the orbit
$\beta$ (including the principal part and all the end-point
corrections) in the $j$th prefragment relative to that in the
non-truncated spherical cavity $A_\beta^{\rm(sph)}$ with the same
radius $R_j$.  The values of $w_\beta$ are found to be similar among
all the important POs, and let us just replace them with
$w_{31}^{(j)}$ for the most important triangle orbit.  Then, the
contributions of POs in the $j$th prefragment can be approximately
given by
\begin{align}
\delta E_j(N)&=\sum_\beta \frac{\hbar^2}{T_\beta^2}A_\beta^{(j)}
\cos\left(S_\beta(e_F;R_j)/\hbar-\pi\mu_\beta/2\right) \NN
 &\approx w_{31}^{(j)}\sum_\beta
 \frac{\hbar^2}{T_\beta^2}A_\beta^{\rm(sph)}
 \cos\left(S_\beta(e_F;R_j)/\hbar-\pi\mu_\beta/2\right) \NN
 &=w_{31}^{(j)}\delta E^{\rm(sph)}(N_j,R_j),
\end{align}
where the prefragment particle number $N_j$ is related to the total
particle number by
\begin{equation}
N_j(e_F)\approx \left(\frac{R_j}{R_0}\right)^3 N(e_F).
\end{equation}
Thus, the shell energy of the 3QS cavity can be approximated by the
sum of shell energies in the spherical cavity $\delta E^{\rm(sph)}$ as
\begin{equation}
\delta E(N)\approx \sum_{j=1,2} w_{31}^{(j)}\delta E^{\rm(sph)}(N_j;R_j).
\label{eq:tf_ito_sph}
\end{equation}
Since the spherical cavity model has magic numbers $N=\cdots, 34, 58,
92, 138, \cdots$, (see e.g., Fig.~2.4 of \cite{RSBook}) the system
will achieve shell energy gain when the prefragment particle numbers
coincide with those magic numbers.

\begin{figure}
\includegraphics[width=\linewidth]{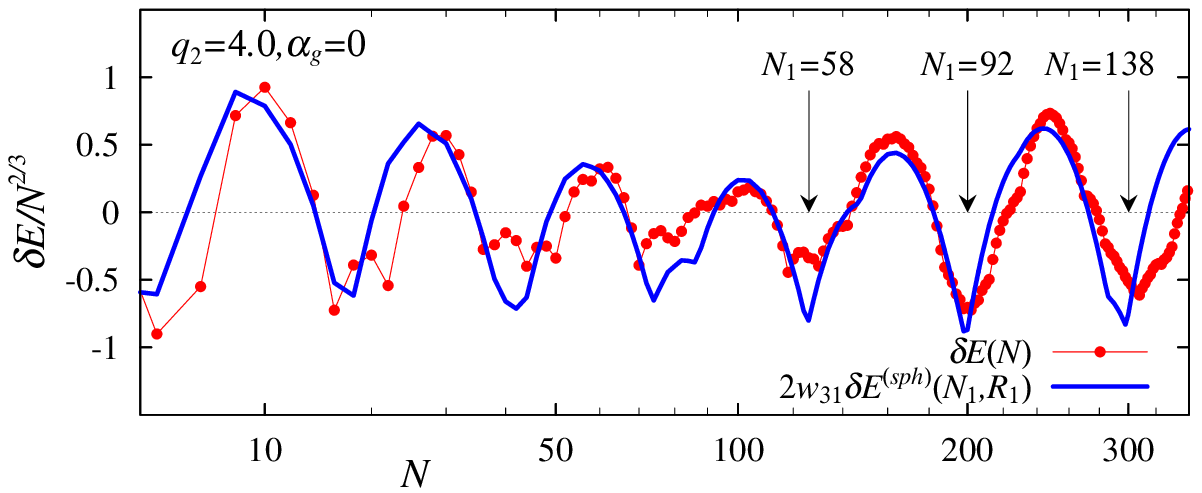} %{scexn_sig.eps}
\caption{\label{fig:sce_sig}
Shell energy of the symmetric 3QS cavity model with the elongation
parameter $q_2=4.0$ as a function of the particle number $N$.  The
dots represent the exact quantum result and the thick solid curves
represent the sum of shell energies of the spherical cavities.
Prefragment magic numbers $N_1(=N_2)$ are indicated by the arrows.}
\medskip

\includegraphics[width=\linewidth]{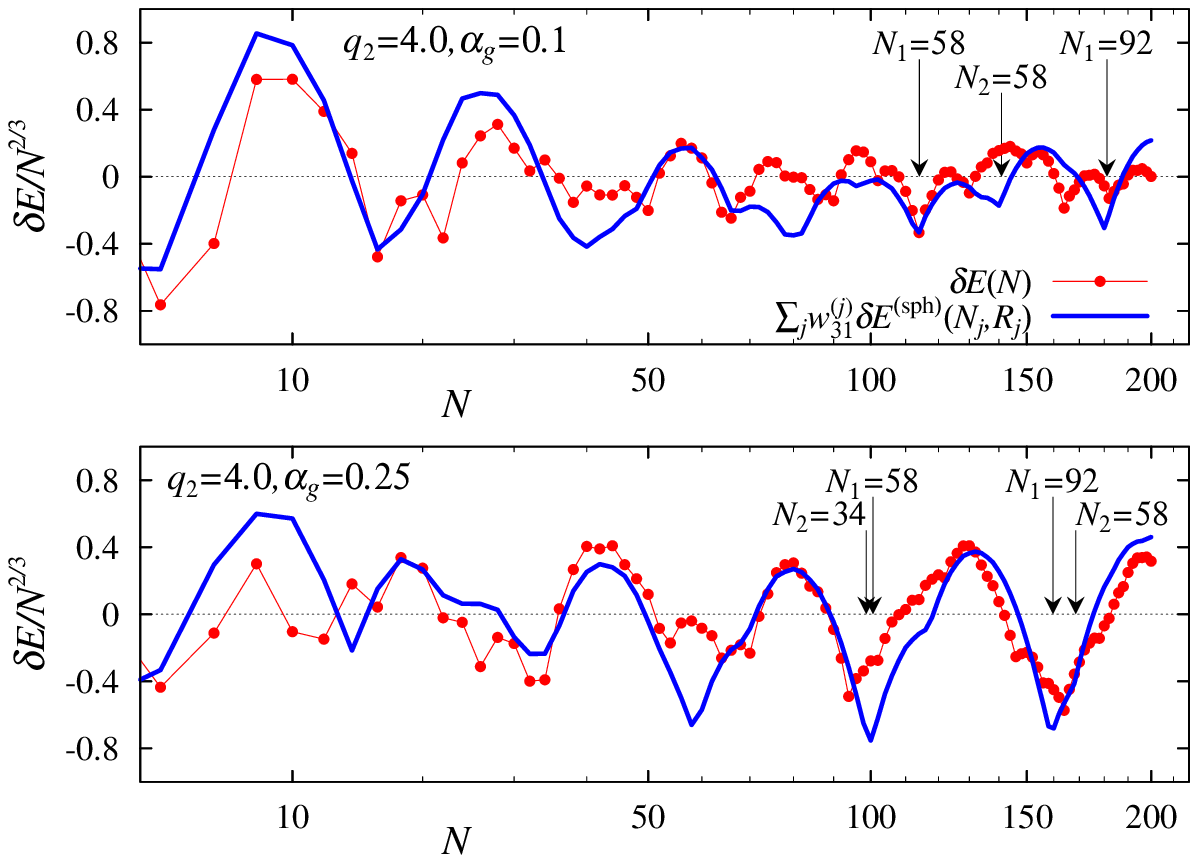} %{scexn_alp.eps}
\caption{\label{fig:sce_alp}
Same as Fig.~\ref{fig:sce_sig} but for asymmetric shapes.
The upper panel is for the elongation parameter $q_2=4.0$ and the
asymmetry parameter $\alpha_g=0.1$.  The lower panel is for the same
elongation parameter but larger asymmetry parameter $\alpha_g=0.25$.}
\end{figure}

Figures~\ref{fig:sce_sig} and \ref{fig:sce_alp} compare the exact
shell energies with the sum of spherical ones given on the right-hand
side of Eq.~(\ref{eq:tf_ito_sph}).  For symmetric shapes shown in
Fig.~\ref{fig:sce_sig}, one sees nice agreement between the two
results, and the shell energy minima are corresponding to the
prefragment magic numbers.  For asymmetric shapes, the agreement
between the two results is not as good as the symmetric case, but one
clearly sees the effect of prefragment magic numbers.  For
$\alpha_g=0.25$, magic numbers of two prefragments make constructive
contributions to give the deep shell energy minima, e.g., around
$N=100$ and $160$.

\begin{figure}
\includegraphics[width=\linewidth]{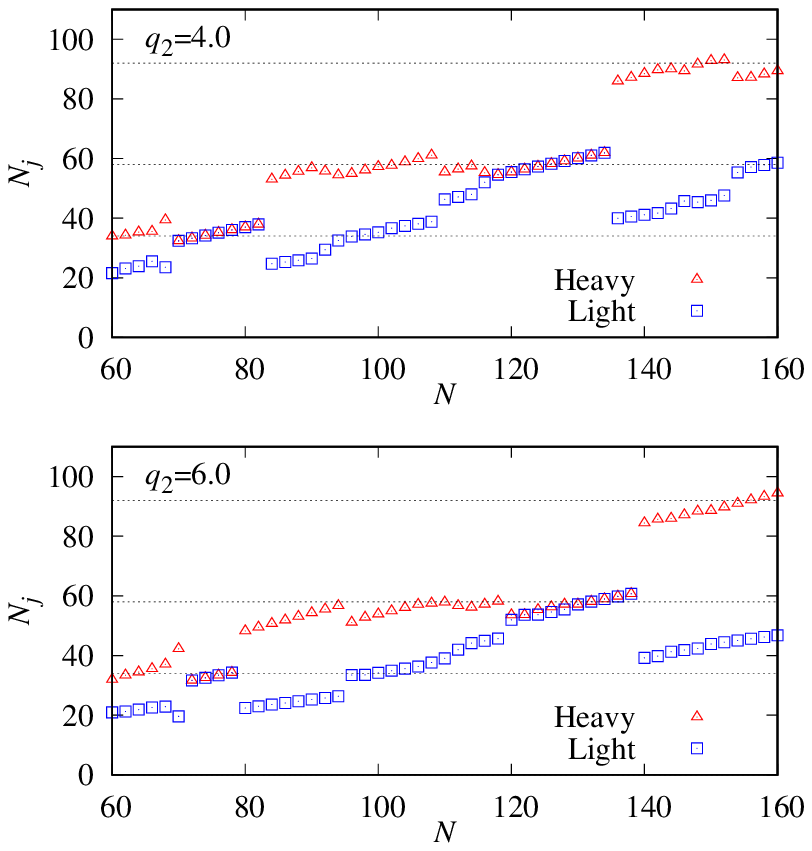} %{abund.eps}
\caption{\label{fig:abund}
Prefragment particle numbers at optimized asymmetries for fixed value
of the elongation $q_2$.  The upper panel is for $q_2=4.0$ around the
saddle, and the lower panel is for the larger elongation $q_2=6.0$.}
\end{figure}

Let us consider the effect of these prefragment magic numbers to the
asymmetric fission.  We calculated the shell energy $\delta E(N)$ as a
function of $q_2$ and $\alpha_g$, and find the value of $\alpha_g$
which minimizes the shell energy for each fixed value of $q_2$.  At
those values of the asymmetry parameter, prefragment particle numbers
are calculated and plotted in Fig.~\ref{fig:abund}.  The horizontal
dotted lines indicate the spherical magic numbers.  It is found that
the heavier prefragment particle number sticks to one of the magic
number and jumps to the next magic number in a stepwise manner with
increasing total particle number $N$.  The result for $q_2=6.0$ is
almost the same as that for $q_2=4.0$, and these prefragment particle
numbers successfully explain the behavior of the experimental data for
the fragment mass distributions shown in Fig.~\ref{fig:fmd}.

\begin{figure}
\includegraphics[width=\linewidth]{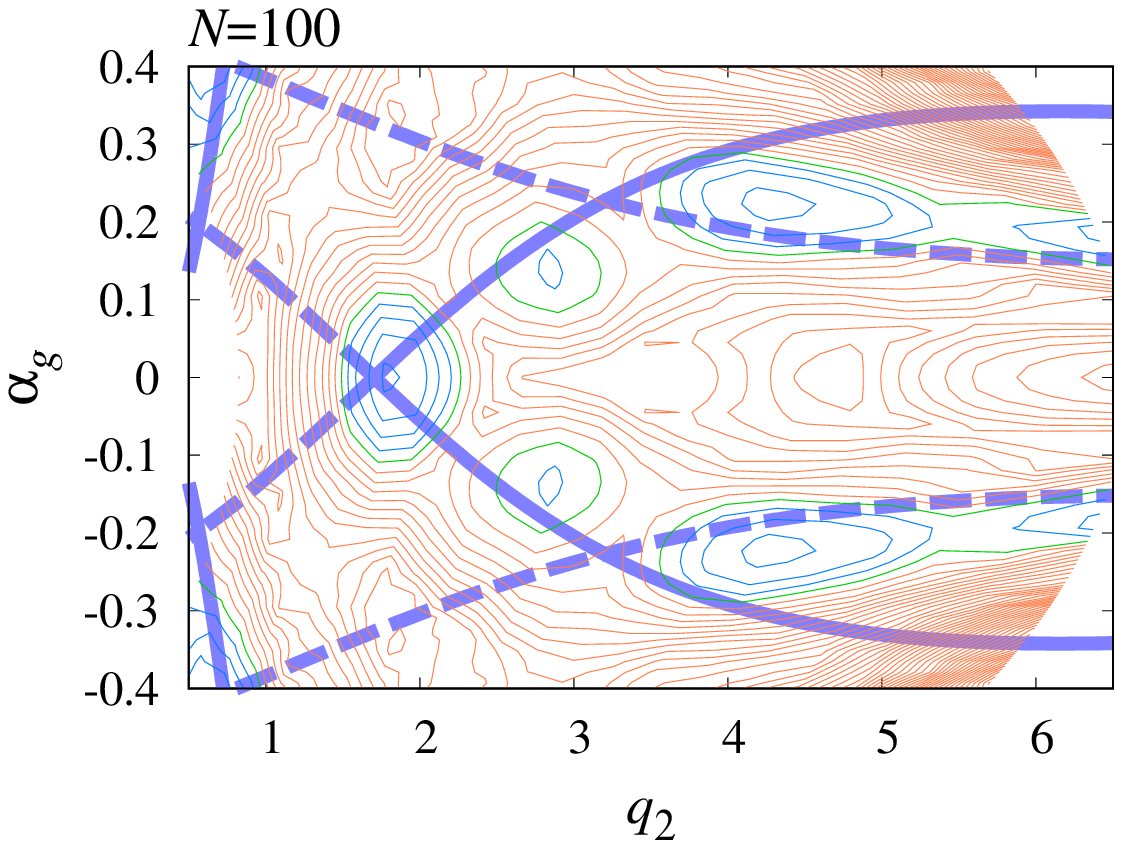} \\ %{sce150.eps}\\
\includegraphics[width=\linewidth]{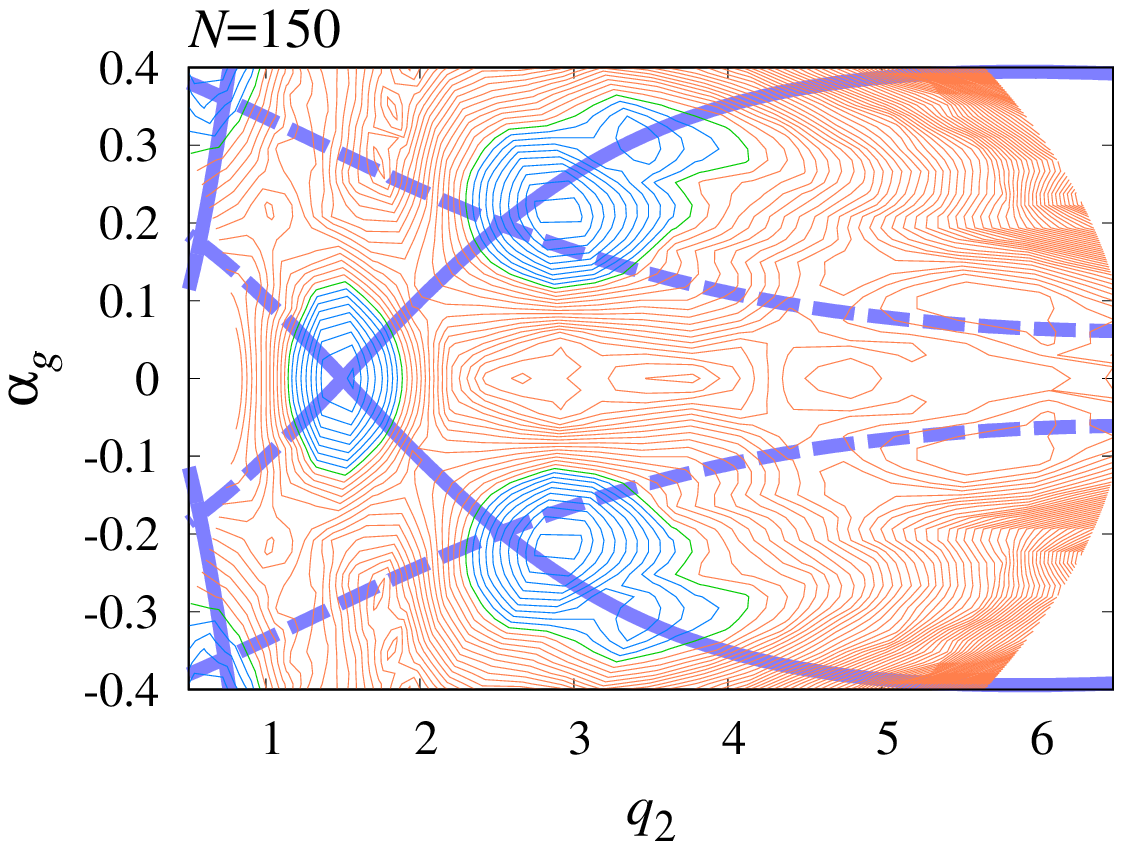}    %{sce100.eps}
\caption{\label{fig:cac}
Contour maps of the shell energies in the shape parameter space
$(q_2,\alpha_g)$ for particle numbers $N=100$ and $150$, that are
chosen for the proton and neutron numbers in actinide region.  Solid
(in blue) and broken (in red) contour lines represent the negative and
positive shell energy, respectively. The pale thick solid and broken
curves represent the constant-action curves (\ref{eq:curve}) for the
prefragment triangle orbits in heavier and lighter prefragments,
respectively, where the prefragment particle numbers take the
spherical magic numbers.}
\end{figure}

Figure~\ref{fig:cac} shows the contour plots of the shell energy for
several particle numbers in the deformation space $(q_2,\alpha_g)$.
One sees valleys running through the strongly elongated asymmetric
shapes.  This curve approximately corresponds to the constant-action
curve of the triangle orbits
\begin{gather}
k_F(N)L_{31}^{(i)}(q_2,\alpha_g)-\frac{\pi}{2}\mu_{31}=(2n+1)\pi,
\NN
L_{31}^{(i)}(q_2,\alpha_g)=\frac{(2n+1+\mu_{31}/2)\pi}{k_F(N)},
\quad (n=0,1,2,\cdots) \label{eq:curve}
\end{gather}
where contribution of triangle orbit to the shell energy
(\ref{eq:sce}) takes minima.  It approximately gives the
condition for the prefragment particle number to coincide with
the spherical magic number.

\begin{figure}
\includegraphics[width=\linewidth]{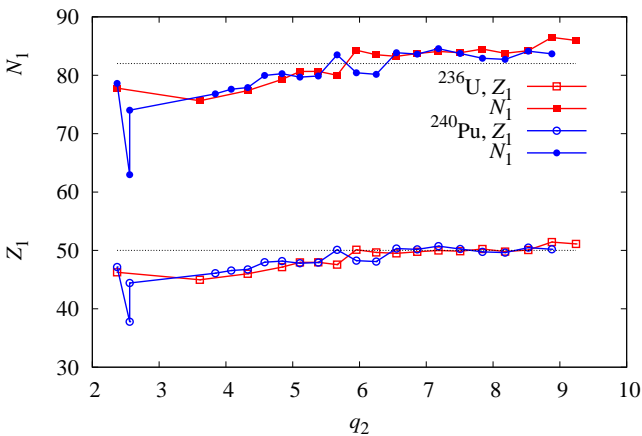}
\caption{\label{fig:nfrag_real}
Prefragment proton and neutron numbers along the fission paths of
\nuc{U}{236} and \nuc{Pu}{240} obtained by the realistic
macroscopic-microscopic calculations\cite{Ichikawa2012}.
The horizontal lines indicate
the spherical magic numbers $50$ and $82$.}
\end{figure}

To see the role of the prefragment shell effect in more realistic
situations, we examined the results of the realistic
macroscopic-microscopic calculations carried out by one of the authors
(T. I.) and his collaborators\cite{Ichikawa2012}.  The potential
energy surface was calculated as a function of five shape parameters
in the 3QS parametrization, and then the fission path was determined
by the immersion method.  Looking at the shapes at the minima and
saddles along the fission paths for \nuc{U}{236} and \nuc{Pu}{240}, we
noticed that the heavier prefragment remains spherical and its radius
$R_1$ is approximately constant.  In Fig.~\ref{fig:nfrag_real}, the
prefragment proton and neutron numbers evaluated by
\begin{equation}
Z_1=\left(\frac{R_1}{R_0}\right)^{1/3}Z, \quad
N_1=\left(\frac{R_1}{R_0}\right)^{1/3}N.
\end{equation}
are plotted against the elongation parameter $q_2$.  One sees that
both proton and neutron numbers stick to the magic numbers $Z_1=50$
and $N_1=82$.  This strongly suggests the significance of prefragment
shell effect to determine the shapes of the nucleus along the fission
path in realistic situations.

In the realistic calculation for \nuc{Hg}{180}, both of the
prefragments are deformed in the fission process.  We expect that our
semiclassical prescription will also be useful in investigating
the role of the prefragment shell effect on the asymmetric fissions
in this region.

\section{Summary}
\label{sec:summary}

In this work, we investigated the shell structures in fission
processes with the 3QS cavity model.  Using the POT, prefragment shell
effect is evaluated as the contributions of POs confined in each of
the prefragments.  As the nuclear body is elongated, neck
configuration develops and the prefragment triangle orbit family makes
a dominant contribution to the shell effect.  The energy valleys are
formed along the constant-action curves where the contribution of the
triangle orbit takes minima.  Since the spherical magic numbers are
approximately given by the action condition of the triangle orbit
family, one has significant prefragment shell effect along the above
constant-action curves, and they are playing significant roles in
determining the fission path on the potential energy surface.

In the present study, the prefragments are fixed at spherical shapes
for simplicity.  This successfully reproduces the experimental
features of the fissions in actinide nuclei where the spherical shell
effect of the heavier prefragments are significant.  However, the
lighter prefragments are usually deformed and the prefragment
deformation should be taken into account for more extensive
description of the fission processes.  This should be critical in
analyzing the asymmetric fission of other mass regions, where both of
the prefragments are expected to be deformed.  In recent realistic
mean-field calculations, the importance of the octupole shape degree
of freedom for the prefragments was suggested\cite{Scamps18,Scamps19}.
It would be an interesting future subject to consider which kinds of
shape degrees of freedom to be taken into account to describe the
optimum fission path.  When the octupole degree of freedom is taken
into account, the effect of local symmetry restorations associated
with the PO bifurcations might play some important
roles\cite{Sugita1998}.

\end{document}